\documentclass[runningheads]{llncs}
\usepackage{graphicx}
\usepackage{cite}
\usepackage{amsmath }
\usepackage{multicol}
\usepackage{tabularx} 

\usepackage{multirow}
 
\usepackage{multirow}
\usepackage[colorinlistoftodos]{todonotes}
\usepackage{mathptmx}
\usepackage{soul}\setuldepth{article}
 \usepackage{float}
\usepackage{amsmath}
\usepackage{cases}
\usepackage{array}

\usepackage{graphicx,changepage}
\usepackage{hyperref}
\hypersetup{ colorlinks=true,
    linkcolor=blue,
    filecolor=magenta,
    citecolor=blue,
    urlcolor=cyan,
    pdftitle={Overleaf Example},
    pdfpagemode=FullScreen,
    }
\usepackage{graphicx}
\graphicspath{ {./images/} }
\usepackage[T1]{fontenc}
\usepackage{geometry}
\usepackage{textcomp}
\usepackage{rotating}

\begin{document}
\title{Knowledge Distillation for Adaptive MRI Prostate Segmentation Based on Limit-Trained Multi-Teacher Models}
\titlerunning{AKD\_MRI}
%
\author{Eddardaa Ben Loussaief\inst{1} \and Hatem Rashwan\inst{1} \and Mohammed Ayad \inst{2} \and Mohammed Zakaria Hassan\inst{1}
\and
Domenec Puig\inst{1}}

\authorrunning{Eddardaa et al.}
%
\institute{Department of Computer Engineering and Mathematics, Universitat Rovira I Virgili. \\
\email{eddardaa.benloussaief@urv.cat} \and Department of Electronic, Electric and Automatic Engineering, Universitat Rovira I Virgili, \\ 43007 Tarragona, Spain
}

 \maketitle              
\begin{abstract}
With numerous medical tasks, the performance of deep models has recently experienced considerable improvements. These models are often adept learners. Yet, their intricate architectural design and high computational complexity make deploying them in clinical settings challenging, particularly with devices with limited resources. To deal with this issue, Knowledge Distillation (KD) has been proposed as a compression method and an acceleration technology. KD is an efficient learning strategy that can transfer knowledge from a burdensome model (i.e., teacher model) to a lightweight model (i.e., student model). Hence we can obtain a compact model with low parameters with preserving the teacher's performance. Therefore, we develop a KD-based deep model for prostate MRI segmentation in this work by combining features-based distillation with Kullback–Leibler divergence, Lovasz, and Dice losses. We further demonstrate its effectiveness by applying two compression procedures: 1) distilling knowledge to a student model from a single well-trained teacher, and 2) since most of the medical applications have a small dataset, we train multiple teachers that each one trained with a small set of images to learn an adaptive student model as close to the teachers as possible considering the desired accuracy and fast inference time. Extensive experiments were conducted on a public multi-site prostate tumor dataset, showing that the proposed adaptation KD strategy improves the dice similarity score by $9\%$, outperforming all tested well-established baseline models.
\end{abstract}
\section{Introduction}

Magnetic Resonance Imaging (MRI) has been widely used for medical imaging diagnosis, i.e., semantic segmentation that consists of pixel-level interpretation by providing; as a result,  so-called segmentation masks. However, medical imaging segmentation remains challenging due to various factors, i.e., limited data resources and image acquisition (imaging modalities and scanning protocols). Deep learning (DL) networks have proven their efficiency in various medical tasks, especially semantic segmentation.  For instance, UNet \cite{r10} is the most used DL architecture for medical imaging segmentation that can handle small training data.
Recently, many efforts have been proposed to upgrade the UNet architecture, i.e., by adopting a strong features extractor, such as ResNet \cite{r21},  as well as by integrating attention blocks or upgrading UNet to support 3D data as described in 3D-UNet \cite{r12}, thereby increasing  computational components and expansion of  the storage. Thus, deploying UNet and its variants in a real-time clinical setting is still challenging. In response to the issues mentioned previously, some recent works have started exploiting lightweight models, i.e., ESPNet \cite{r20}, ENet \cite{r19}, MobileNet-v2 \cite{r22}, etc., in real-time medical imaging segmentation, but with accepting the degrade of accuracy. Thus, there is a dilemma in balancing the model performance and the low computational cost. 

Hinton et al. \cite{r8} have proposed a new learning schema, knowledge distillation (KD), to overcome the above drawbacks. KD tends to distill the learning from a well-trained teacher to a lightweight student model, enhancing the latter's performance by preserving the teacher's performance. KD has recently gained the attention of several researchers in semantic segmentation. For instance, Xu et al.\cite{r9} proposed using a growing teacher assistant network (GTAN) for CT liver segmentation. The role of the proposed GTAN is to leverage the difference between the size of the models used, i.e., teacher and student.
Furthermore, Li et al. \cite{r26} adopted mutual KD to improve cross-modality segmentation, transferring prior knowledge from CT images to MRI images. In \cite{r7}, the knowledge adaptation for brain segmentation adopts the same teacher and student models network and exploits the coordinate distillation (CD) that incorporates the channel and space information. Additionally, in \cite{r3}, KD is applied for 3D  optical microscope image segmentation. However, the aforementioned methods teach the student network by only distilling the logits on the final teacher's output. Thus, the student still yields segmentation results with low accuracy.

Therefore, many recent methods developed feature-based KD methods to consider the intermediate features generated during the learning process and distill more discriminated information. In \cite{r1}, He et al. introduced an affinity distillation module to optimize the similarity between teacher and student features. In turn, Liu et al. \cite{r2} adopted two distillation schemes by applying pixel and structured pair-wise distillation based on intermediate feature learning. Qin et al.\cite{r4} investigated a new module called importance maps distillation that aims to match the student's feature maps with the teacher's feature maps through a re-scaling process. Due to the complex structure of medical data and the difference in the scanning protocols and patient privacy, the aforementioned methods may not yield good segmentation results in real-time clinic use.

To this end, Huang et al. \cite{r6} proposed to collect data from different hospitals to train an adaptive teacher model. To deal with the information disclosure and sharing of patients' data.
Training a teacher model with images from different sources will improve the model's generalization and it could yield good segmentation results. However, it can also introduce biases and confounding factors that negatively affect the model's performance. For example, hospitals may use different imaging machines with different settings, resulting in image quality and resolution variations. The model may learn to rely on these variations to distinguish between different classes, which can lead to poor generalization when applied to new data \cite{r34}. Here,  Multi-teacher KD can potentially help to address the issue of training DL models with medical images from different sites and hospitals with different scanning protocols. In the case of medical image segmentation, multi-teacher KD can help capture a wider range of features and patterns from the data, improving the student model's ability to generalize and adapt across different sources and protocols. The different teacher models may be trained on different subsets of the data, allowing the student model to learn from a diverse set of examples \cite{r35}. Additionally, medical datasets are always limited to a small set of images. Accordingly, we need to multi-teacher KD, but teachers trained with limited data. As far as we know, almost no methods have considered multi-teaching distillation training with limited data to build a valuable system-based distillation for medical imaging segmentation.

Consequently, in this work, we aim to develop multi-teacher framework-based distillation to tackle the problem of medical data sharing and limitations. We propose a novel learning scheme where the student model learns from the cross-teacher's logits and tends to mimic the middle teachers' features during the distillation process. Concretely, we perform two distillation architectures: 1) We adopt the intermediate features-based mapping along with KL\_divergence as logits-based to transfer the knowledge from the off-the-shelf MRI segmentation models to a compact model, i.e. a lightweight student model. 2) The majority of multi-teacher based distillation vote to the teacher that has the highest logits distribution or perform the averaging of the soft targets. Unlike these simple distillation manners, we propose an adaptive KD in that our strategy guides the training of the student model by aggregating the knowledge from $N$ teachers (we adopt here $n=3$)  through an adaptive weight. Furthermore, our selection of the teachers' networks no longer only considers the the-off-shelf segmentation networks with a high number of parameters, but also we share the dataset between the teachers. Specifically, to tackle the patient disclosure information and to respond to the real-time segmentation for the clinical setting, we split our data into three small sets to train the teachers separately. We conduct extensive experiments to validate the proposed distillation schemes on public prostate MRI datasets collected from different sources \cite{r17}. The results demonstrate the remarkable improvement of the lightweight model achieving an improvement of $~9\%$  for the MRI prostate tumor segmentation dice score compared to the state-of-the-art.

\section{Methodology}
\subsection{Knowledge Distillation (KD)}
This work is DL-based KD technique for domain adaptation, it aims to generate a lightweight model through the guidance of a large and complex teacher network. During the distillation process, the student model imitates the teacher to gain a competitive performance. Thus, we can obtain a compact model KD-trained with a few parameters, and its efficiency is also improved.

\begin{figure}[H]
 \begin{center}
  \includegraphics[width=0.7\textwidth, height = 0.45\columnwidth]{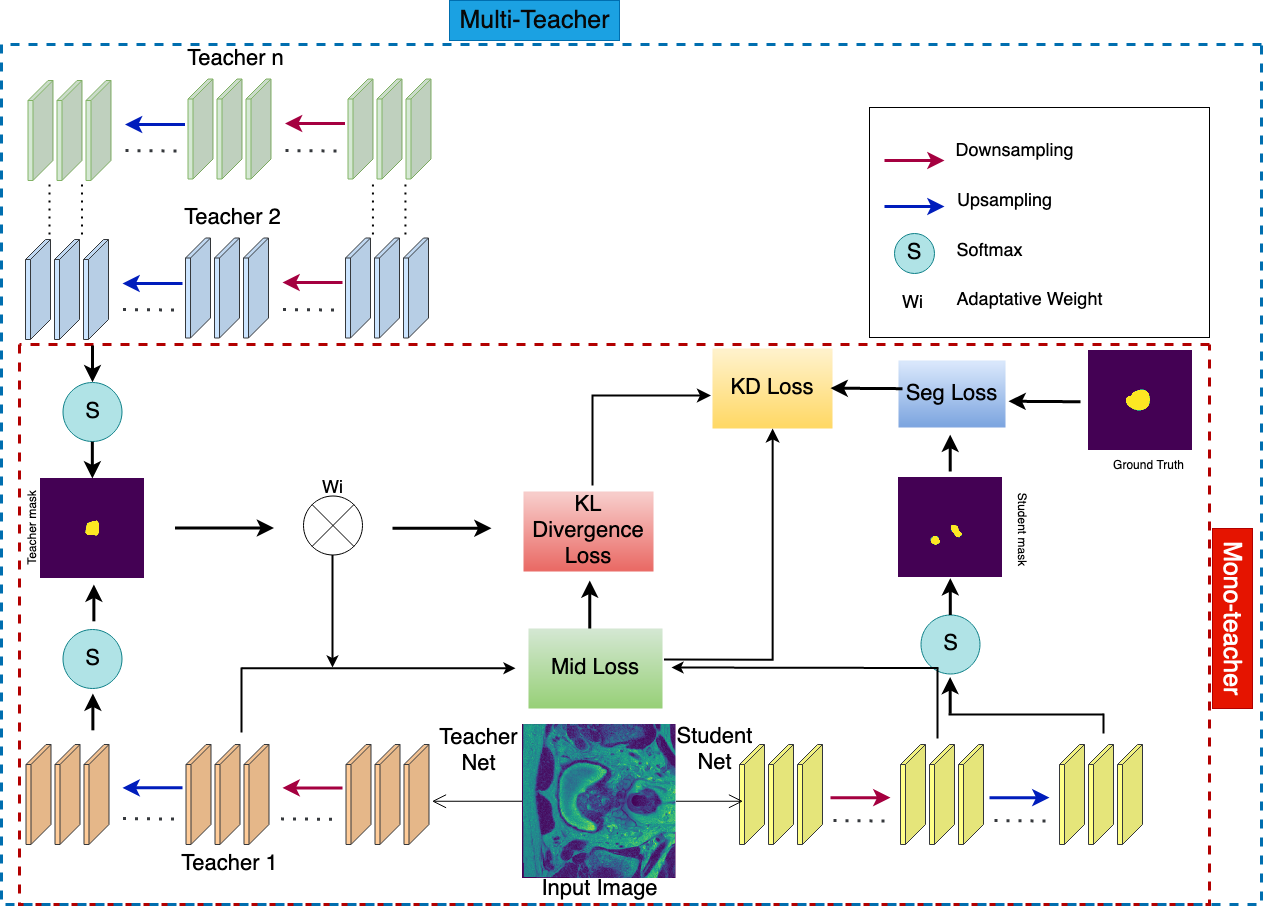}
  \caption{Overview of the proposed  distillation pipeline.}
  \label{fig1}
  \end{center}
  \vspace{-1cm}
\end{figure}

The entire distillation structure shown in Fig. \ref{fig1} comprises two schemes. The first is mono-teacher distillation, where the student learns from a single cumbersome teacher well-trained. The second distillation scenario tends to learn the lightweight model from multi-teacher trained with limited data. We aim to enable the student to learn extra information from multiple teachers and to empower it to be adaptive to real-time prostate MRI segmentation for any hospital dataset to deal with private data sharing. Benefiting from MRI prostate multi-sources \cite{r17} data that we use for evaluating the two KD schemes, we split the large dataset into three small sets and then train the teachers individually using one set. Hence, we enable the ensemble of teachers to distill the student network through transferring process that uses the same losses as a mono-teacher. 
Our model ensures that the teacher models are diverse enough to capture a wide range of data variations while avoiding redundancy that could lead to overfitting. Additionally, we ensure that the student model can effectively learn from the ensemble of teacher models, which may require specialized training techniques based on effective distillation loss functions.

\subsection{Network optimization}
\label{sec1}
 Fig. \ref{fig1} illustrates the distillation pipeline introduced in this paper, where the architecture takes the MRI imaging as a stack of 2D grayscale slices. We adopt a transferring architecture comprised of logits-based and feature-based distillation.\par Logits-based distillation has been conducted in \cite{r8} as an efficient architecture to learn the student network from the soft targets provided by a well-optimized teacher model. In our paper, we propose a distillation architecture that uses soft targets that contains the class probability and the essential information for the student to mimic the output of the pre-trained teacher. The transfer probability is as follows:
\begin{equation}
\centering q_{i}=\frac{exp(z_{i}/\lambda)}{\sum_{j}exp(z_{j}/\lambda)},
\label{eq1}
\end{equation}
where $q_{i}$ indicates the input probability and $z_{i}$ denotes the ith class of the logits output, i.e., $z_{0}$ background and $z_{1}$ prostate. $\lambda$ is a hyperparameter referred to as the temperature to balance the distillation loss. Inspired by \cite{r8},  we determine a prediction  distillation that enables the lightweight model to learn the prediction probability of the final output segmentation of the teacher model, i.e., softmax output. The logits-based distillation loss is precisely measured by Kullback–Leibler (KL) divergence across the student's and teacher's softmax predictions. Thus, prediction distillation loss is defined by: 
\begin{equation}
    \centering KL_{loss}=\frac{1}{N}\sum_{i}^{N}KL(p_{i}^{s}||p_{i}^{t}),
    \label{eq2}
\end{equation}
 where the KL divergence function is denoted by $KL(.)$, where $N=w\times h$, all the pixels in segmentation, and $p_{i}^{s}$ and $p_{i}^{t}$ indicate the probability of ith pixel pair in the segmentation map of the student's and teacher's softmax  outcome, respectively. 
 
 Motivated by \cite{r1, r4}, along with the prediction distillation loss measured in (\ref{eq2}), we explore an intermediate distillation loss to conduct the feature-based distillation that considers an importance distillation loss and an affinity distillation loss to interpolate the student's and teacher's feature maps and optimize the similarity across them. Considering the intermediate distillation loss across middle feature maps for the teacher and student networks, we adopt a $Mid_{loss}$ that measure the importance and affinity losses proposed in \cite{r1,r4}. $Mid_{loss}$ and defined as follows:
\begin{equation}
        Mid_{loss}=\sum_{(i,j) \in P}||\frac{M_{i}^{s}}{||M_{i}^{s}||_{2}} -   \frac{M_{j}^{t}}{||M_{j}^{t}||_{2}}||_{1} + \sum_{(i,j) \in P}||V_{rc}^{s}-V_{rc}^{t}||_{2},
\end{equation}
where $M_{i}^{s}$ and $M_{j}^{t}$ are the importance maps of the ith and jth layer extracted from the student and teacher, respectively. $||.||_{1}$ and $||.||_{2}$ are the L1 and L2 normalization. P is the collection of pixel pair positions of the same size. $V_{rc}^{s}$ and $V_{rc}^{t}$ represent the region contrast vectors for the student and teacher, respectively ( as described in \cite{r1,r4}). The importance loss matches the teacher's feature maps to the student's features maps by re-scaling the latter. However, the region contrast consists of the transfer of the relationship information across networks by enabling the student model to imitate the region contrast between the region information of the foreground area and the background area.


To compare the student's segmentation mask with a hard target, i.e., ground truth, we consider the segmentation loss $Seg_{loss}$ that adopts Lovasz-softmax loss \cite{r28}, which is particularly useful for class imbalance present and boundary alignment, in addition to dice similarity loss \cite{r32}, as follows:
\begin{equation}
    \centering Seg_{loss}= \alpha_{1} Dice_{loss} + \alpha_{2} Lovasz_{loss},
    \label{eq3}
\end{equation}
where $\alpha_{1}$ and $\alpha_{2}$ are hyperparameters set to 0.2 and 0.3, respectively.  

As shown in Fig. \ref{fig1}, to obtain an end-to-end trainable student network, we exploit a combination of losses to perform the global distillation loss, $KD_{loss}$. Hence the total distillation for the mono-teacher scenario is given by the following formula:
    \begin{equation}
        \centering KD_{loss}=Seg_{loss} +\alpha Mid_{loss} + \beta KL_{loss},
        \label{eq4}
    \end{equation}
where $Seg_{loss}$ and $KL_{loss}$ presented in (\ref{eq2}) and (\ref{eq3}), respectively. $\alpha$ and $\beta$ are set to 0.1. The hyperparameters setting was settled after such empirical attempts  to conduct efficient segmentation results.\par 
For multi-teacher distillation, we introduce an adaptive strategy to empower the student to learn from all the teachers in the prediction distillation process. Thus, we propose an adaptive weight to compute the final output of three teacher models, $p_{i}^{t}=w_{j}\sum_{j}p^{t_{j}}$, where $j = 1:n$ and $n$ is the number of teachers, (in this work $n=3$ and $w_{j} \in \{w_{1}, w_{2}, w_{3}\}$). The adaptive weights $w_{j}$ calculated by:

\begin{math}
\centering
  \left\{
    \begin{array}{c}
      w_{1}=Dice_{loss}(p^{t_{1}},y)/\sum_{j}(Dice_{loss}(p^{t_{j}},y)\\
  w_{2}=Dice_{loss}(p^{t_{2}},y)/\sum_{j}(Dice_{loss}(p^{t_{j}},y)). \\
  w_{3}=Dice_{loss}(p^{t_{3}},y)/\sum_{j}(Dice_{loss}(p^{t_{j}},y))
    \end{array}
  \right.
\end{math} \\
\\
The new $KL_{loss}$ measures the prediction distillation between the final  $p_{i}^{t}$ and the student's output, $p_{i}^{s}$ as shown (\ref{eq2}). We also calculate the $Mid_loss$ across the three teachers and the student network as follows:
\begin{equation}
    Mid_{loss}= w_1Mid_{loss}(T1,S) + w_2Mid_{loss}(T2,s) + w_3Mid_{loss}(T3,S).
\end{equation}

\section{Experiments}
\subsection{Dataset and evaluation metric}
\label{secdata}
We conducted our experiments on a publicly available MRI prostate dataset \cite{r17}, where the data comprises six different sites that were collected from three public datasets, i.e. The sites 1,2, and 3 are from  NCI-ISBI2013 \cite{r29}, and I2CVB \cite{r30}, respectively. Whilst sites 4, 5, and 6 are collected from PROMISE12 \cite{r31}. The data consists of 116 patient cases with a total slice number of $1740$ with their corresponding segmentation masks. For mono-teacher distillation, we split the data  into $80\%$, $5\%$, and $15\%$ for training, validation, and testing, respectively. According to multi-teacher distillation, we adopt the following site pair partitions: (1,2),  (3,4), and (5,6) for training teachers 1, 2, and 3, respectively. We compute the dice similarity score to assess  MRI segmentation results to different baseline architectures. 

\subsection{Setup}
A series of extensive experiments were conducted to validate the efficiency of our distillation pipeline. We select the powerful off-shelf-segmentation models, such as DeepLabv3+ \cite{r3},  PSPNet \cite{r13}, FCN\_ResNet101 \cite{r14}, and ResNext-101-32x8d \cite{r35} as teacher networks, and several publicity available lightweight networks, i.e.,  ENet \cite{r19}, ESPNet \cite{r20}, and MobileNet-v2 \cite{r22} as student models. We adopted the PyTorch official setup of all tested networks to train and test them on Gefore RTX 3050 ti(8GB). Adam optimizer was used during the training with its default configuration ($\beta_1 0.9$ and $\beta_2=0.999$) and an initial learning rate of 0.01. We also used CyclicLR to schedule the learning rate with the lowest learning rate of 0.000001 and step size=2000. We trained the models to converge with 100 epochs.

\subsection{Results}
\subsubsection{\textbf{\textit{Mono-teacher based distillation:}} }
We aim to learn a student network from a single teacher for the first scenario in our distillation pipeline. The teacher is well trained with the training set of cite{r17}, including the six sites. To demonstrate the power of the distillation scheme in the efficiency enhancement of a lightweight model, we conducted empirical series, where we adopted ENet \cite{r19}, ESPNet \cite{r20}, and MobileNet-v2 \cite{r22} as student networks with a low number of parameters. In turn, we selected complex teacher networks, i.e.,  DeepLabv3+ \cite{r3},  PSPNet \cite{r13}, FCN\_ResNet101 \cite{r14}, and ResNext-101-32x8d \cite{r35} to guide the aforementioned student models. DeepLabv3+ and PSPNet networks, with a backbone of ResNet50 \cite{r21}, and ResNet101 used as a backbone for FCN. The aforementioned networks are tailored for medical imaging tasks in the literature with high accuracy. We present quantitative segmentation results for MRI tumor prostate from the dataset in \cite{r17} in the table \ref{tab:single teacher}. As shown, with the appropriate selection of the teacher networks, all students are adept at learning and mimic the teachers' capability and achieve higher performance compared to the level of the teacher's performance. It is obvious to remark that ENet, ESPNet, and MobilENet-v2 embrace the maximal improvement of 5.87\%
(80.03 to 85.9), 9.1\% (75.5 to 84.3) and 6.9\% (78.4 to 85.3) in dice score, respectively. It is clear that all student networks are improved, however, they are still under the teacher's performance, but saving millions of parameters. The best improvement was yielded with ENet and FCN\_ResNet101 as a teacher with a Dice score of $85.9\%$. In Fig. \ref{fig2}-(column 4), we visualize the segmentation masks of two test samples where the lightweight network MobileNet-v2 was distilled from PSPNet. As shown, there is a significant difference across the segmentation masks given by MobileNet-v2 as baseline (column 3) and after distillation concretely for the second sample.
\begin{table}[ht]
\caption{CROSS EXPERIMENTS RESULTS BETWEEN A SINGLE TEACHER AND STUDENT ON MRI PROSTATE TUMOR SEGMENTATION. }
\begin{center}
\begin{tabular}{c |c|c|c}
    \hline
  Method & Dice score ($\% \pm std$)  & Params (M) & FLOPs (G)\\ \hline \hline
   \multicolumn{4}{c}{Teacher Networks (Baseline)}\\ \hline\hline
   
    T1: DeepLabV3+ & 89.01 $\pm 0.001$ & 56.8  &273.94   \\
                               T2: Fcn\_ResNet101 & 90.0 $\pm 0.002$ & 51.96  & 199.74   \\
                                T3:PSPNet & 86.3 $\pm 0.004$ &46.71   & 207.7   \\
                                T4:ResNext101 & 85.02 $\pm 0.006$ & 86.74  &96.54   \\
                               
                               \hline
    \hline
    \multicolumn{4}{c}{Student Networks}\\ \hline\hline
     S1: ENet Baseline & 80.03 $\pm 0.007$ & 353.76k  & 2.2   \\
     T1\_KD\_ENet & 84.6 $\pm 0.002$ & --&-- \\
     T2\_KD\_ENet &85.9 $\pm 0.001$&--&--\\
     T3\_KD\_ENet &84.2 $\pm 0.001$&--&-- \\
     T4\_KD\_ENet &83.7 $\pm 0.004$&--&--
\\     \hline \hline
   S2: ESPNet Baseline & 75.2 $\pm 0.081$ & 183.72k  & 1.27   \\
   T1\_KD\_ESPNet &81.9 $\pm 0.001$&--&-- \\
   T2\_KD\_ESPNet &84.3$ \pm 0.002$&--&--\\
   T3\_KD\_ESPNet&81.7 $\pm 0.004$&--&--\\
   T4\_KD\_ESPNet&80.5 $\pm 0.006$&--&--\\

   \hline\hline
    S3: MobileNet-v2 Baseline & 78.4 $\pm 0.008$ &2.23  & 19.84  \\
    T1\_KD\_MobileNet-v2 &82.3 $\pm 0.003$&--&--\\
    T2\_KD\_MobileNet-v2&81.2 $ \pm 0.001$&--&--\\
    T3\_KD\_MobileNet-v2 &85.3 $\pm 0.001$&--&--\\
    T4\_KD\_MobileNet-v2 &80.1 $\pm 0.006$&--&--\\
    \hline \hline
  
\end{tabular}
\end{center}
\label{tab:single teacher}
\vspace{-1cm}
\end{table}
The FLOPS and number of parameters are calculated using the Flops counter in PyTorch framework \textbf{ptflops} by feeding the input size, $1\times384\times384$, 
with the trained model to interpret the computational complexity of this latter as listed in the table \ref{tab:single teacher}. The ESPNet provided the lowest FLOPs among all tested models with FLOPs of $1.27 G$. The distillation with a well-trained teacher significantly improved the segmentation results with a huge reduction in FLOPs and number parameters.

\subsubsection{\textbf{\textit{Multi-teacher based distillation :}}}
We performed experiments to explore the multi-teacher distillation concept and verify if the student could learn more information through distilling from n teachers trained with limited data. In this work, we selected three teachers, T1: Deeplabv3+ \cite{r3},T2: PSPNet \cite{r13}, and T3: FCN-ResNet101 \cite{r14}, which yielded the best dice score as  baselines. Our multi-teacher network selection was also based on the model that empowers the lightweight student to gain the highest enhancement, i.e., we adopted PSPNet, DeepLabv3+, and FCN\_ResNet101 to conduct multi-teacher distillation architecture. The three teachers were trained with limited data; each teacher trained with two sites out of the six sites of the whole dataset as described in Section \ref{secdata}. As described in Section \ref{sec1}, we introduced an adaptive weight determinate by the dice loss for each teacher separately for the prediction distillation loss and the intermediate loss. Table \ref{tab: multi-teacher} lists the performance of the student networks distilling from three powerful teachers. Referring to table \ref{tab:single teacher}, all the students, i.e., ENet \cite{r19}, ESPNet \cite{r20}, and MobileNet-v2 \cite{r22} grasp a significant improvement in the dice score, when they distilled from Deeplabv3+ \cite{r3}, PSPNet \cite{r13}, and FCN\_ResNet101 \cite{r14} already tailored on state\_of\_the\_art medical imaging segmentation. 
\begin{table}[H]
\caption{CROSS EXPERIMENTS RESULTS BETWEEN A THREE TEACHER AND STUDENT ON MRI PROSTATE TUMOR SEGMENTATION. }
\begin{center}
\begin{tabular}{c |c}
    \hline
  Method & Dice score ($\% \pm std$)  \\ \hline \hline
   \multicolumn{2}{c}{Multi-Teacher Distillation }\\ \hline\hline

     S1: ENet Baseline & 80.03 $\pm 0.007$  \\
     T1+T2+T3\_KD\_ENet & 83.8 $\pm 0.004$  \\
     
   \hline \hline
   S2: ESPNet Baseline & 75.2 $\pm 0.081$  \\
   T1+T2+T3\_KD\_ESPNet &83.4 $\pm 0.003$ \\

   \hline\hline
    S3: MobileNet-v2 Baseline & 78.4 $\pm 0.008$ \\
    T1+T2+T3\_KD\_MobileNet-v2 &84.8 $\pm 0.004$\\
    
    \hline \hline
  
\end{tabular}
\end{center}
\label{tab: multi-teacher}

\end{table}
As shown in Table \ref{tab: multi-teacher}, all the student networks achieved better segmentation results, where they gain an improvement of up to 3.77\% (80.03 to 83.8), 8.2\% (75.2 to 83.4), and 6.4\% (78.4 to 84.8) for ENet, ESPNet, and MobileNet-v2, respectively. ESPNet is the lightest model with a high dice score compared to the ENet with a slight difference of a Dice score of $1\%$. Furthermore, we can observe that the multi-teacher distillation scheme gained a comparable dice score to the mono-teacher distillation. However, our aim is not only to construct a distillation pipeline that enables a lightweight model to imitate the strong teacher's performance, as obtained from mono-teacher distillation but also to build a compact adaptive model that can cope with multi-source data and data disclosure problems, as well as can be used on limited devices without extra computation cost. In other words, we are not seeking to empower the student to achieve higher performance from multiple teachers than a single teacher till we emphasize that distilling from multi-teacher could help the student to benefit from each teacher's capability separately.

\begin{figure}[H]
 \begin{center}
  \includegraphics[width=0.6\textwidth, height=7.5cm]{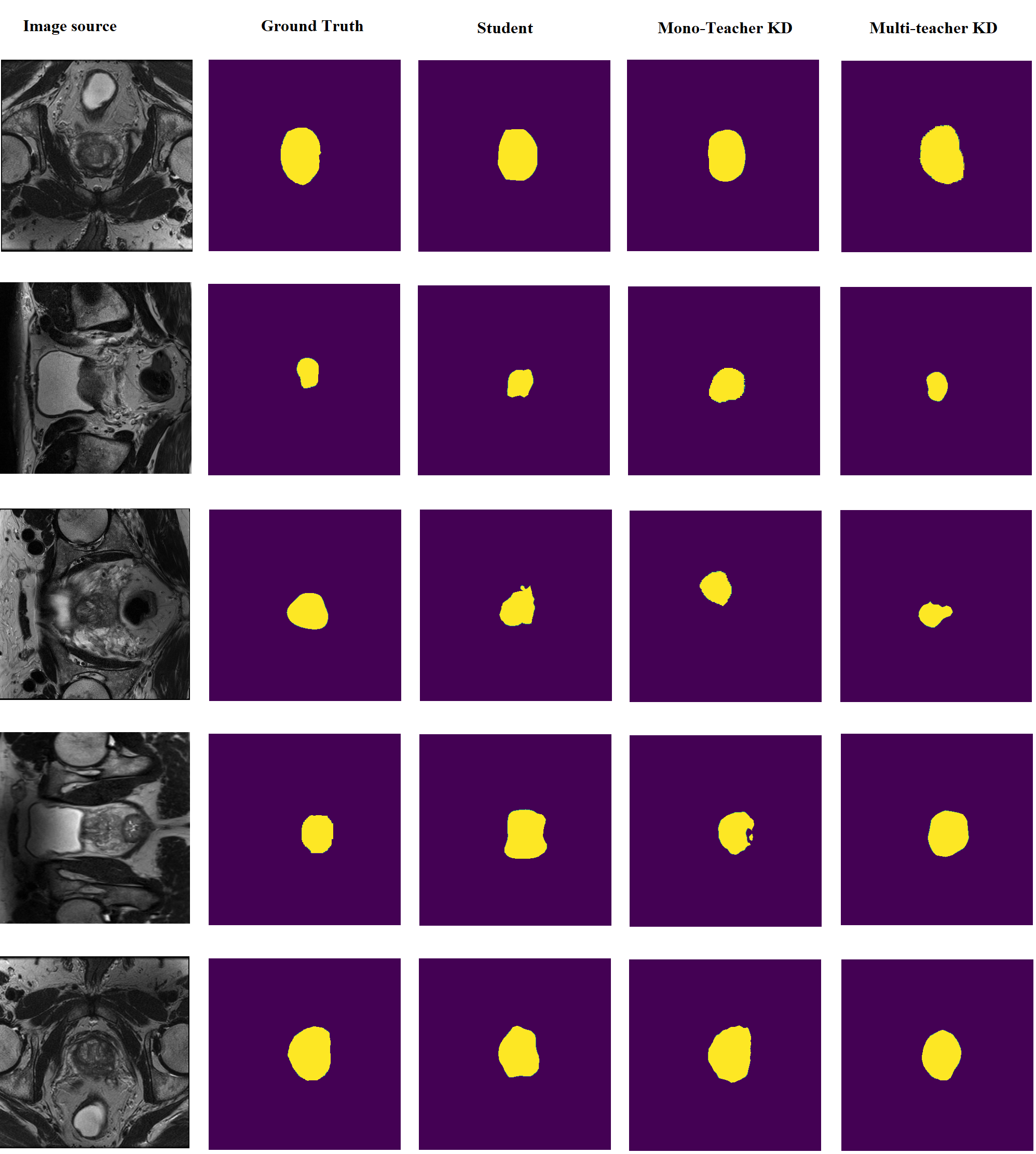}
  \caption{Visualization of the prediction results of knowledge distillation architecture.}
  \label{fig2}
  \end{center}
  \vspace{-1cm}
\end{figure}

Qualitatively, in Fig. \ref{fig2}, the segmentation results from multi-teacher are visualized to demonstrate that their predictions are more discriminated and representative of that from a single teacher in the same cases (see rows 1 and 2 in Fig. \ref{fig2}). In turn, the third row represents the case where the mono-teacher distillation produces a better segmentation result than the multi-teacher. Thus, we validate that the paradigm of multi-teacher distillation could perform well for medical imaging analysis, i.e., segmentation and classification. We can say that the student networks can improve even with teachers trained with a small dataset. Yet, referring to the obtained result, the multi-teaching scheme should be improved to leverage the prediction results, and this could be our coming work.
 
\section{Conclusion}

To sum up, This work presented two knowledge distillation schemes based on single and multi-teacher. This study has shown that our proposed approach of applying multi-teacher framework-based distillation can overcome the problem of medical data sharing and data limitation while maintaining the model's performance with as low as possible computational costs and outperforming the state-of-the-art in terms of dice score for MRI prostate segmentation by an improvement of $~9.0\%$ in some cases. The results proved that we could preserve the accuracy of the complex models trained with limited data in lightweight models that can be executed on devices with limited resources. However, our study was only conducted on MRI scans, and we did not address other medical imaging modalities, such as CT scans and X-Ray. Thus, in intermediate work, we aim to apply the distillation scheme to different image modalities. Future work explores the potential use of multi-teaching online distillation, where the teacher and student could be both end-to-end trainable in parallel through an online pipeline to boost the generalization of the student network within a faster inference time.
\section*{Acknowledgement}
This research was done thanks to The RadioCancers project (Grant agreement ID: PID2019-105789RB-I00) funded by the National Spanish Ministry of Science and Innovation, SPAIN.

%
%
%
\bibliographystyle{splncs04}
\bibliography{thebibliography}

\end{document}